\documentclass[aps,prb,twocolumn,floatfix]{revtex4}
\usepackage{psfig}

\begin{document}

\title{
\parbox{30mm}{\fbox{\rule[1mm]{2mm}{-2mm}\Large\bf\sf PREPRINT}}
\hspace*{4mm}
\parbox{100mm}{\footnotesize\sf
Submitted for publication in {\it Optics Letters} (2005) }\hfill
\\[5mm]
Coupling into the slow light mode in slab-type photonic crystal waveguides}

\date{\today}

\author{Yurii~A.\ Vlasov and Sharee~J.\ McNab}
\address {IBM T.~J.\ Watson Research Center, Yorktown Heights, NY 10598, USA}

\begin{abstract}
Coupling external light signals into a photonic crystal (PhC) waveguide becomes
increasingly inefficient as the group velocity of the waveguiding mode slows down. We
have systematically studied the efficiency of coupling in the slow light regime for
samples with different truncations of the photonic lattice at the coupling interface
between a strip waveguide and a PhC waveguide. An inverse power law dependence is found
to best fit the experimental scaling of the coupling loss on the group index. Coupling
efficiency is significantly improved up to group indices of 100 for a truncation of the
lattice that favors the appearance of photonic surface states at the coupling interface
in resonance with the slow light mode.\end{abstract}

\maketitle

\noindent

Planar two-dimensional (2D) slab-type photonic crystals (PhC) have attracted much
attention recently as a possible platform for densely integrated photonic circuits.
Engineering of the photonic dispersion utilized in planar devices might provide unique
functionalities for integrated photonics unattainable with conventional approaches. For
example slowing down the propagation velocity of light \cite{ref1,ref2} in PhC waveguides
has been proposed for compact delay lines and all-optical storage devices \cite{ref3}.
However, increase of the group index in the slow light regime prevents efficient coupling
into the PhC waveguide due to increasingly large impedance mismatch. Several recent
theoretical studies \cite{ref7,ref8,ref9} indicate that the exact termination of the
photonic lattice at the coupling interface is important for improving impedance matching.
It has also been suggested that surface states localized at the PhC interface can play a
significant role in the coupling process \cite{ref4,ref5,ref6}.

In this Letter we study experimentally the dependence of the coupling efficiency on the
group index for different terminations of the photonic lattice. Further we explore the
possibility to improve mode matching and obtain efficient coupling by tuning the photonic
surface states at the PhC interface in resonance with the slow light mode.

To experimentally study the coupling efficiency, PhC structures were fabricated on a
silicon-on insulator 200mm wafer with 1$\mu$m BOX layer on a standard CMOS fabrication
line as described elsewhere \cite{ref10,ref11}. PhCs with a triangular lattice of period
$a$=437nm were defined by etching holes with radius $R$=109nm through a silicon layer
with thickness $d$=225nm. PhC waveguides were formed by omitting one row of holes (W1
waveguide) in the lattice along the $\Gamma$-K direction. In order to probe the influence
of surface termination, a set of samples was fabricated in which the truncation of the
PhC waveguides at the strip/PhC interface was varied by changing the termination
parameter from $\tau$=0 to $\tau$=1 as shown in the inset of Fig.1. The length of the PhC
W1 waveguides $L$ was kept approximately constant with 22 full unit cells (~10$\mu$m).
Light from a broadband source (four coupled LEDs with 50nm linewidth each) was coupled in
and out of the photonic chip through polymer-based inverted fiber couplers using tapered
and micro-lensed PM fibers \cite{ref10,ref11}. High resolution spectra were also measured
with a tunable diode laser having a 100MHz linewidth. Access strip waveguides with
460x220nm cross-section are butt-coupled to the PhC W1 waveguides through a lateral
taper\cite{ref10} with the final width of 757nm corresponding to $\sqrt{3}a$.
Transmission spectra from the PhC waveguide circuits were normalized on the transmission
spectrum of a strip waveguide circuit without a PhC. Owing to small side-wall surface
roughness with standard deviation $~$1.5nm as measured with an AFM, the propagation loss
in analogous strip waveguide and PhC waveguide has been measured recently to be as low as
8$\pm$2db/cm and 5$\pm$0.2db/cm at 1650nm \cite{ref10,ref11}, correspondingly.

\begin{figure}[tb]
\begin{center}
\leavevmode \psfig{figure=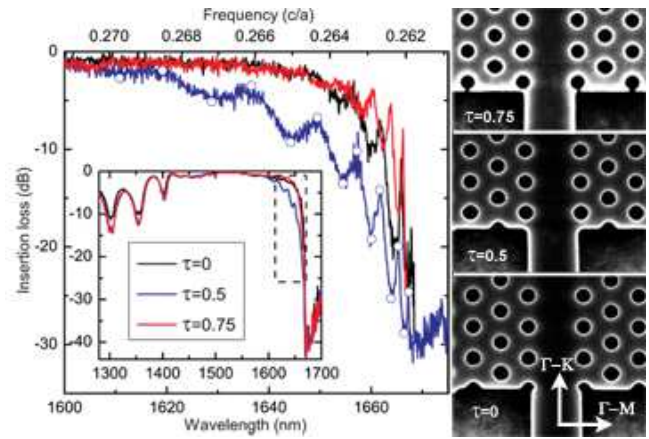,width=85mm}
\end{center}
\caption{Inset: Set of low spectral resolution (5nm) transmission spectra of samples with
different termination of the strip/PhC interface. Black, blue and red corresponds to
terminations $\tau$=0, 0.5, and 0.75, correspondingly. The blown-up portion of the
spectra measured with 60pm resolution is shown on the main graph. Open blue dots show the
position of maxima and minima of oscillations. Right panel shows the SEM images of the
structures investigated.}
\label{fig1}
\end{figure}

The inset of Fig.1 presents a set of transmission spectra recorded with optical spectrum
analyzer for TE polarized light for wavelengths ranging from 1300 to 1700nm. As is seen,
the spectra for samples with terminations $\tau$=0, $\tau$=0.5, and $\tau$=0.75 are
almost identical for most of the wavelengths. Spectra exhibit a sharp cutoff at 1670nm
corresponding to the onset of the W1 waveguiding mode \cite{ref11,ref12}. It is seen that
at wavelengths around 1600nm coupling at strip/PhC interface is almost perfect
0.3$\pm$0.1dB for all the terminations. This excellent coupling is not surprising since
the width of the access strip waveguide is chosen to match closely both the geometrical
spread of the mode in the PhC waveguide and its group index far from the slow light
regime \cite{ref10,ref11,ref13}.

At wavelengths longer than approximately 1600nm and closer to the waveguide onset cutoff
at 1670nm a notable difference in the spectra is observed. This region corresponds to
where the wavevectors $\textbf{k}$ approach the Brillouin zone edge, and is characterized
by an increasingly slow group velocity. Figure 1 shows the same set of transmission
spectra for this wavelength range measured with an LED source with a spectral resolution
of 60pm. Strong Fabry-Perot oscillations, especially noticeable for the spectrum of the
sample with $\tau$=0.5, are observed indicating large reflections at the coupling
interface. The distance between minima and maxima $\Delta$$\lambda$ of the oscillations
is decreasing from ~10nm to below 1nm towards the mode onset cut-off indicative of an
increasingly small group velocity.  The spectral positions of the maxima and minima of
the oscillations can be used to extract the spectral dependence of the group index as
$n_g$=$\lambda^{2}$$/$(4·$L$·$\Delta$$\lambda$). Group indices approaching 100 are
typical for the last visible maxima around 1667nm. It is also seen that the amplitudes
of the maxima $I_{max}$ and minima $I_{min}$ are gradually decreasing toward the cut-off,
while $V$, the fringe visibility $V$=($I_{max}$-$I_{min}$)$/$$(I_{max}$+$I_{min}$), is
actually increasing toward the cutoff approaching 0.8 for the last fringes around 1667nm.
The latter indicates that the fringe amplitude is not seriously affected by the low
coherence of the LED source (50nm line width). This is confirmed by comparison with
spectra measured with high-coherence tunable laser shown in the inset of Fig.2.

\begin{figure}[tb]
\begin{center}
\leavevmode \psfig{figure=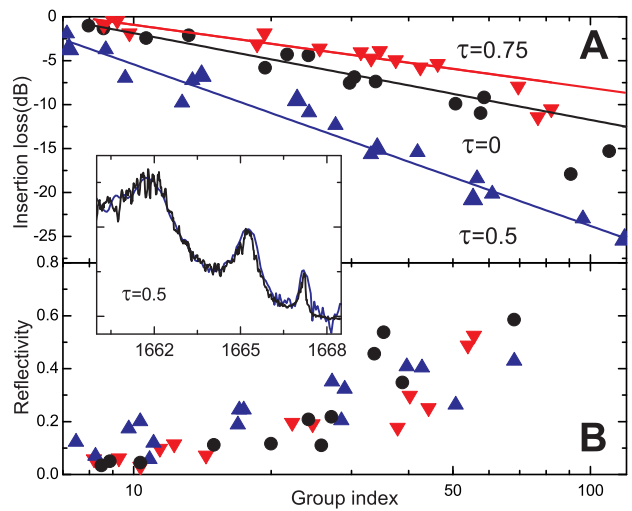,width=85mm}
\end{center}
\caption{Inset: Transmission spectra of a sample with $\tau$=0.5 measured with LED (60pm
resolution, blue curve) and laser (20pm steps, black curve) as a source. A). Coupling
losses of the pair of strip/PhC interfaces measured for different terminations. B).
Reflectivity at the strip/PhC interface for different terminations. Black, blue and red
symbols correspond to terminations $\tau$=0, 0.5, and 0.75, correspondingly. Lines show
results of fitting with an inverse power law with the exponent -0.99, -1.84, and -0.72
for $\tau$=0, 0.5, and 0.75, respectively.}
\label{fig2}
\end{figure}

Comparison of fringe maxima in the spectra for different terminations in Fig.1 implies
that contribution of propagation losses is negligible. Indeed the fringe maxima differ by
almost 10dB in the spectra  (for example for $\tau$=0 and $\tau$=0.5), while the length
of the PhC waveguide $L$ is identical (note the identical spectral positions of maxima
and minima). If transmission losses in the slow light regime were the dominant source of
loss we would expect fringe amplitudes measured at the same wavelength to be identical
for different terminations, the opposite of what we observe experimentally.

The evident difference in the amplitude of the maxima for samples with different
termination indicates that the main source of damping is increasingly inefficient
coupling at the PhC/strip interface. If we assume that the propagation losses inside the
PhC are negligible the amplitude of the maxima corresponds to the combined coupling
losses at the input and output PhC/strip interfaces. Figure 2a presents a log-log plot of
the coupling losses (amplitude of fringe maxima) as a function of the group index for PhC
waveguides with different termination. The reflectivity $R$ of the PhC/strip interface
can also be extracted from the fringe visibility $V$=$2R/[1+R^{2}]$ assuming that the
reflectivities of the input and output interfaces are equal. The dependence of the
interface reflectivity $R$ on the group index is shown in Fig.2b. Three different samples
were measured for each termination.

\begin{figure}[tb]
\begin{center}
\leavevmode \psfig{figure=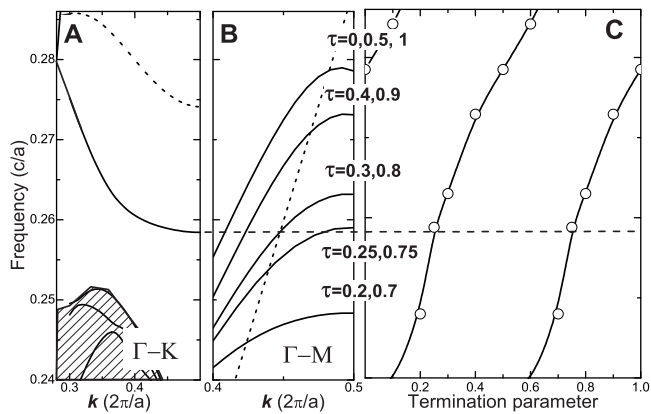,width=85mm}
\end{center}
\caption{A). Projected photonic band diagram of the W1 PhC waveguide calculated with a 3D
planewave method along $\Gamma$-K direction for $R/a$=0.25 and slab thickness 0.52$a$.
The solid line correspond to the z-even TE-like mode of interest. Dashed line is z-odd
TE-like mode. Dotted line depicts the light line. B). Photonic band diagrams along
$\Gamma$-M direction of the PhC slab truncated at various terminations $\tau$. C).
Frequency position of the surface mode at $\textbf{k}$=0.5 for different terminations
$\tau$.}
\label{fig3}
\end{figure}

Although visible even in Fig.1 the differences between different terminations become
evident analyzing Fig.2a. Here the experimental dependence of coupling efficiency on the
termination is best fitted with inverse power law dependence, which gives the exponent
-0.99 and -1.84 for $\tau$=0 and 0.5, respectively. The best coupling though is provided
by the termination $\tau$=0.75 where the fitting gives exponent as -0.72. Surprisingly
there is no noticeable dependence of interface reflectivity on termination as seen in
Fig.2b. To the best of our knowledge these are the first experimental measurements of
both coupling and reflectivity of the PhC interface in the slow light regime.

Several recent publications examined theoretically the coupling efficiency of the
strip/PhC interface for different terminations of the lattice \cite{ref7,ref8,ref9}.
Although the slow light regime was intentionally omitted from consideration, for
frequencies far from the mode cut-off it has been shown that terminations around $\tau$=0
are preferred over $\tau$=0.5. It has been argued \cite{ref8} that this is a result of
better impedance matching due to the spatial periodicity of the PhC Bloch wave and,
correspondingly, impedance at the termination. Indeed our experimental results clearly
show that $\tau$=0 termination has far superior coupling than $\tau$=0.5 even for group
indices as high as 100. However the best performance is experimentally observed for
$\tau$=0.75. To better understand this result the photonic band diagrams of a PhC W1
waveguide (Fig.3a) and a PhC slab truncated along $\Gamma$-M direction (see Fig.3b) were
calculated with the 3D planewave expansion method \cite{ref13}. It is seen from Fig.3
that the surface states appear in the photonic gap with dispersion (Fig.3b) and spectral
position (Fig.3c) depending strongly on the termination parameter. The truncations
$\tau$=0, 0.5 and $\tau$=1 correspond to surface states tuned to frequencies much higher
than the PhC mode. Surface states do not contribute to propagation at the strip/PhC
interface and these terminations are equivalent in this respect. Terminations around
$\tau$=0.25 and $\tau$=0.75, however are characterized by surface states tuned almost in
resonance with the PhC waveguide slow light mode. Moreover the surface state dispersion
is almost flat at these frequencies reflecting the strong localization of the surface
mode within only a few periods from the interface. Correspondingly not only is the
impedance at the PhC termination strongly modified by the presence of surface states, but
the group indices of the PhC slow light mode and surface states are also nearly matched.
Based on these observations we can argue that experimentally measured performance
indicate that surface states do play a significant role in coupling.

In conclusion we have experimentally measured coupling efficiency and reflectivity at the
strip/PhC interface in the slow light regime. A strong inverse power law dependence of
the coupling efficiency on group index was found for different terminations of the PhC
lattice at the interface. Experimental results and theoretical calculations suggest that
terminations with photonic surface states tuned in resonance with the PhC slow light mode
provide the best coupling efficiency. This finding can shed light on many other coupling
phenomena in PhC like, for example, recently discovered beaming and focusing of light
exiting the 2D PhC waveguide structure \cite{ref5,ref6}.

Fruitful discussions with D.N.Chigrin and A.Y.Petrov are gratefully acknowledged. This
work was supported in part by DARPA Slow light project(J.Lowell, DSO), grant
N00014-04-C-0455.

\end{document}